\documentclass{article}

\usepackage[margin=1in]{geometry} 

\usepackage{siunitx}
\usepackage{amsmath}
\usepackage{algorithm}
\usepackage{bbold}
\usepackage{graphicx}
\usepackage{subcaption}
\usepackage{titling}
\usepackage[noblocks]{authblk} 

\preauthor{\begin{flushleft}\large \lineskip 0.5em \Authfont}
\postauthor{\par\end{flushleft}}

\renewcommand\Authfont{\large\bfseries}     
 
\usepackage{hyperref}
\hypersetup{
    colorlinks=true,
    allcolors=blue
}

\title{\Large\bfseries Methods for adjusting for covariate measurement error in flexible modelling of functional form: results of a blinded, controlled neutral comparison simulation study}

\author{Mohammed Sedki PhD\thanks{Equally contributing authors.}}
\author{Aris Perperoglou PhD\protect\footnotemark}
\author{Anne C. M. Thi\'ebaut PhD}
\author{Steve Ferreira Guerra}
\author{Paul Gustafson}
\author{Frank E. Harrell, Jr} 
\author{Willi Sauerbrei} 
\author{Michal Abrahamowicz}
\author{Laurence S. Freedman}

\author[ ]{\protect\\ \vspace{0.5em} \large on behalf of the international STRengthening Analytical Thinking for Observational Studies (STRATOS) Initiative}

\affil[1]{Universit\'e Paris-Saclay, UVSQ, Inserm, Gustave Roussy, CESP, Oncostat, Villejuif, France}
\affil[2]{Quantitative Sciences Innovation, GSK, London, UK}
\affil[3]{Department of Epidemiology, Biostatistics and Occupational Health, McGill University, Montreal, Canada}
\affil[4]{Department of Statistics, University of British Columbia, Vancouver, Canada}
\affil[5]{Department of Biostatistics, Vanderbilt University School of Medicine, Nashville, Tennessee, USA}
\affil[6]{Institute of Medical Biometry and Statistics, Faculty of Medicine and Medical Center - University of Freiburg, Freiburg, Germany}
\affil[7]{Information Management Services Inc., Calverton, Maryland, USA}

\date{\small \today}

\begin{document}
\maketitle

\begin{abstract}
Covariate measurement error is pervasive in epidemiological research and distorts estimated exposure–outcome associations, yet correction methods have been studied almost exclusively under linear modelling assumptions. Their behaviour when the underlying association is non-linear—and is itself estimated with flexible regression, remains poorly characterised. We report a blinded, multi-stage neutral comparison simulation study, conducted within the STRATOS initiative, evaluating measurement error correction coupled with flexible modelling of functional form. Six families of correction methods (pointwise and coefficient-based Simulation Extrapolation [SIMEX], Bayesian inference on the logit and risk scales, Multiple Imputation [MI], and Regression Calibration [RC]) were each combined with B-splines (BS), penalised splines (PS), fractional polynomials (FP), and natural splines (NS), yielding 23 analytic methods. Methods were applied to case–control data generated under five functional forms (J-shape, linear, two threshold models, and saturation) across simulated datasets spanning varying sample sizes, replication substudy sizes, error magnitudes, and error distributions, with classical additive error and a replication substudy for error calibration. Performance was assessed by the log mean squared error of the estimated function over the central 95\% of the exposure distribution. Pointwise SIMEX was the most accurate and most robust approach overall, followed by Bayesian methods and RC when paired with PS, FP, or NS; MI performed less well, and Bayesian estimation with unpenalised BS performed worst. PS, FP, and NS were near-equivalent, whereas BS was consistently inferior. No single method dominated across all scenarios, underscoring the value of sensitivity analyses.

\bigskip
\textbf{Keywords:} \emph{Measurement Error}, \emph{Non-Linear Effects}, \emph{SIMEX}, \emph{Multiple Imputation}, \emph{Regression Calibration}, \emph{Bayes}, \emph{Splines}, \emph{Fractional Polynomials}, \emph{neutral comparison design}.
\end{abstract}

\section{Introduction}
\label{sec:introduction}
Measurement error is a frequent, but often neglected, issue in medical research and epidemiological studies. It can stem from various sources, including inaccuracies of measurement instruments, errors during data coding, self-reporting, or using single measurements for variables that change over time. The increasing use of data not originally collected for research, such as routine healthcare data, will increase the prevalence of measurement error. Despite its known presence, measurement error in covariates (like exposure and confounder variables) is often not addressed in published research. A systematic review of high-impact journals in 2016 found that while 44\% of original research publications mentioned measurement error, only a small fraction (7\% of those, or 3\% overall) used methods to investigate or correct for it~\cite{brakenhoff18, shaw18}. This makes it challenging for readers to determine the robustness of reported findings.

It is widely accepted that measurement error in covariates introduces bias and imprecision in estimates of exposure-outcome relationships and the consequences of ignoring measurement error when fitting regression models are well-documented in~\cite{keogh20} and \cite{carroll06}. It leads to bias in reported relationships, diminished power, and masked data features~\cite{carroll06}. The direction and magnitude of the bias can be unpredictable, particularly when multiple covariates are measured with error or when errors are systematic or differential with respect to the outcome. Special statistical methods are required to account for measurement error, and their implementation requires information about the type and size of the error.

To correct the bias, many adjustment methods have been developed, including: Regression Calibration (RC)~\cite{rosner89}, Simulation Extrapolation (SIMEX)~\cite{cook94}, Bayesian methods~\cite{gustafson04, gustafson05}, Multiple Imputation (MI)~\cite{cole06, freedman08}, Moment Reconstruction (MR)~\cite{freedman08}, Moment-Adjusted Imputation (MAI)~\cite{thomas11}, and likelihood-based methods~\cite{keogh14, shaw20}. These methods have been extensively studied and applied~\cite{keogh14}, often focusing on linear regression models or generalised linear models. However, the relationship between the covariate and outcome has usually been assumed to be linear.

Yet, the complexities of dealing with measurement error could be amplified when the true relationship between the covariate measured with error and the outcome is not linear, such as in dose-response studies. In such cases, the functional form of the relationship is unknown, necessitating the use of flexible modelling techniques, like splines~\cite{wood17, perperoglou19} or Fractional Polynomials~\cite{royston94, sauerbrei06}. However, measurement error in the covariate can distort the identification of this true non-linear functional form resulting in bias in the estimated non-linear function and misleading inferences about the nature of the relationship~\cite{keogh20}. Thus, we are left with a gap in understanding how to adjust for covariate measurement error when using flexible modelling techniques to capture non-linear relationships.

To address this gap and provide guidance for applied researchers, the STRATOS (STrengthening Analytical Thinking for Observational Studies) initiative~\cite{sauerbrei14}, through a collaboration between Topic Group 2 (Selection of Variables and Functional Forms) and Topic Group 4 (Measurement Error and Misclassification), initiated a simulation project. The overarching aim of this joint project was to evaluate and compare different methods for estimating the true, potentially non-linear, relationship between an outcome variable ($Y$) and a continuous covariate ($X$) when $X$ is measured with error. The project employed a blinded, multi-stage neutral simulation study design to rigorously compare methodological approaches~\cite{thiebaut25}. It involved a Data Generation and Evaluation team and three distinct Methods teams applying approaches based on Bayesian methods, MI and RC and SIMEX. Each of these methods was implemented in conjunction with four flexible modelling techniques: B-splines (BS), P-splines (PS), Fractional Polynomials (FP), and Natural Splines (NS)~\cite{hastie09, wood17, perperoglou19}. More details about the design may be found in~\cite{thiebaut25}.

We present here results from this project. The paper is organised as follows: Section~\ref{sec:background} presents notation and the concepts of measurement error and nonlinearity in regression. Section~\ref{sec:datagen} describes in detail the simulation experiment that was performed. Section~\ref{sec:results} details the results comparing the performance of the different methods, and there is a discussion in Section~\ref{sec:discussion}.

\section{Background: Measurement Error and Non-Linearity in Regression}
\label{sec:background}

\subsection*{Fundamental Concepts}

Consider a regression setting where the objective is to model the relationship between an outcome variable $Y$ and a continuous covariate $X$, which we will refer to as an exposure variable. The fundamental challenge arises when $X$ cannot be observed directly. Instead, an error-prone measurement is observed, generalised as $X^\ast$, which is a contaminated version of the true value $X$.

The most studied error structure is the classical measurement error (CME) model, where the observed exposure $X^\ast$ is the sum of the true, unobserved variable $X$ and an independent, additive error term $U$, such that $X^\ast = X + U$. The error $U$ is assumed to have a mean of zero and be independent of the true exposure $X$ and also to be independent of the outcome variable $Y$.

Simply substituting the observed $X^\ast$ for the true $X$ in the regression model leads to a distortion of the estimated exposure-outcome relationship. In the case of a simple linear regression of $Y$ on a single covariate measured with classical error, the estimated regression coefficient is biased towards the null, a phenomenon known as attenuation. However, in multivariable settings or when the relationship between $X$ and $Y$ is non-linear, the direction and magnitude of the bias are less predictable and the estimated coefficient may be further from zero than is the true coefficient. Other effects of substituting $X^\ast$ for $X$ in the regression model are reduced power and masked data features~\cite{carroll06}. In combination, these different effects render inferences unreliable and potentially lead to incorrect scientific conclusions. Furthermore, there might be cases where the assumption of a linear relationship between a continuous covariate and an outcome is a convenient but biologically implausible simplification. The true nature of the dose-response relationship, represented by the function $f(X)$ in the general regression model $\mathbb{E}(Y\mid X)$, is frequently unknown and may exhibit substantial non-linearity as illustrated in subsection~\ref{part:functions}. Correctly identifying the functional form of $f(X)$ is paramount, as mischaracterisation can lead to flawed inferences about risk thresholds, optimal exposure levels, and the overall impact of the exposure.

To overcome the rigidity of linear assumptions, a range of flexible regression techniques has been developed. Methods such as BS, NS, PS and FP allow the data to dictate the shape of the relationship between the covariate and the outcome. These approaches model $f(X)$ as a flexible, smooth curve, often providing a more faithful representation of the underlying biological process. The central challenge investigated in this paper is that the accurate estimation of this non-linear function is severely compromised when the covariate $X$ is measured with error, as the error-prone data $X^\ast$ can lead to distorted estimates of the shape of $f(X)$ itself. While the consequences of classical measurement error in linear models are well-characterised, its impact on estimated non-linear relationships is substantially more complex. Simulation work has confirmed that measurement error generally attenuates and linearises the true non-linear exposure-outcome relationship, a finding consistent with earlier investigations in survival analysis~\cite{keogh12}. However, further research also reveals that the impact of measurement error is far from simple or uniform. The extent of the distortion depends on a complex interplay between the magnitude of the error, the sample size, the true functional form, and the underlying distribution of the true covariate~\cite{ferreira26}.

\section{Simulation Study}
\label{sec:datagen}
We organise the description of the study design using the "ADEMP" principles~\cite{morris19}: Aims, Data generation, Estimands, Methods and Performance measures.

\subsection{Aims}
The aim of the study was to evaluate and compare measurement error correction methods coupled with flexible modelling approaches with respect to their accuracy in estimating a variety of exposure ($X$) - response ($Y$) relationships in the presence of measurement error in the exposure.

\subsection{Data generation}
A neutral comparison simulation study was meticulously structured to facilitate a rigorous and unbiased comparison of analytical approaches across a range of plausible scenarios. The study was split into three parts: stage 1 involved a simulation study on a small number of datasets, with information on underlying data characteristics and results blinded to the methods teams. The primary aim of this stage was to allow the methods teams to create the code scripts that would run on a wider set of datasets. Stage 2 involved applying these code scripts to an extended set of data. In stage 3, replications of data sets with the same characteristics as in stage 2 were performed, and natural splines were added in a non-blinded extension to the study. More details on the different stages are given in~\cite{thiebaut25}. The next sub-sections describe the design and parameters of the simulation study.

The foundation of the simulation was the generation of datasets where a binary outcome $Y_i$ for an individual $i$ (where $Y_i = 1$ denotes an event and $Y_i=0$ a no event) was linked to a true, unobserved continuous covariate $X_i$ via an unconditional logistic regression model. The data were structured as a case-control study with four controls per case. The relationship was defined by a potentially non-linear function $f(X_i)$ on the logit scale: $\text{logit}\Big(\mathbb{P}\big(Y_i = 1 \mid X_i\big)\Big) = f(X_i)$. For each individual, an error-prone version of the covariate, $X_i^\ast$, was generated according to an additive CME model: $X^\ast_i = X_i + U_i$. The random error term $U_i$ was characterised by $\mathbb{E}\big(U_i \mid X_i) = 0$ (unbiasedness) and $g(Y_i|X_i,X^*_i)=g(Y_i|X_i)$ (non-differentiality with respect to the outcome). The structure of the error term, specifically its variance $\sigma^2_U$ and its distributional form, was varied across different simulation scenarios, as detailed subsequently. Each main study dataset consisted of $N$ independent observations $\big(Y_i, X^\ast_i\big)$, where $N$ was also varied.

To enable the estimation of parameters related to the measurement error process, a validation substudy of size $n$ was simulated for a randomly selected subset of individuals from the main study. The parameter $n$ was also varied. For these individuals, an independent replicate measurement of $X^\ast_i$, denoted $X^\ast_{i2}$, was generalised. These replicates were drawn from the same measurement error distribution as the $X_{i1}^\ast$ (equivalent to $X_{i}^\ast$ above) observed in the main study, satisfying $X_{i1}^\ast = X_i + U_{i1}$ and $X_{i2}^\ast = X_i + U_{i2}$ where $U_{i1}$ and $U_{i2}$ were independent and identically distributed with $\mathbb{E}\big(U_{i1}\big) = \mathbb{E}\big(U_{i2}\big)=0$ and $\text{Var}\big(U_{i1}\big) = \text{Var}\big(U_{i2}\big) = \sigma^2_U$.

The true underlying functional form $f(X)$, the distribution of the true covariate $X$, the specific parameters of the measurement error model (variance and distribution of $U$), and the true values of $X_i$ were kept hidden from the Methods teams throughout the analysis stages. This blinding protocol helped to ensure an unbiased evaluation and comparison of the analytical methods~\cite{thiebaut25}.

\subsubsection{Simulation Scenarios and Parameters: Stage 1}
\label{part:functions}
\begin{figure*}[htp]
\centering
\begin{minipage}{0.31\textwidth}
  \centering
  \includegraphics[width=\textwidth]{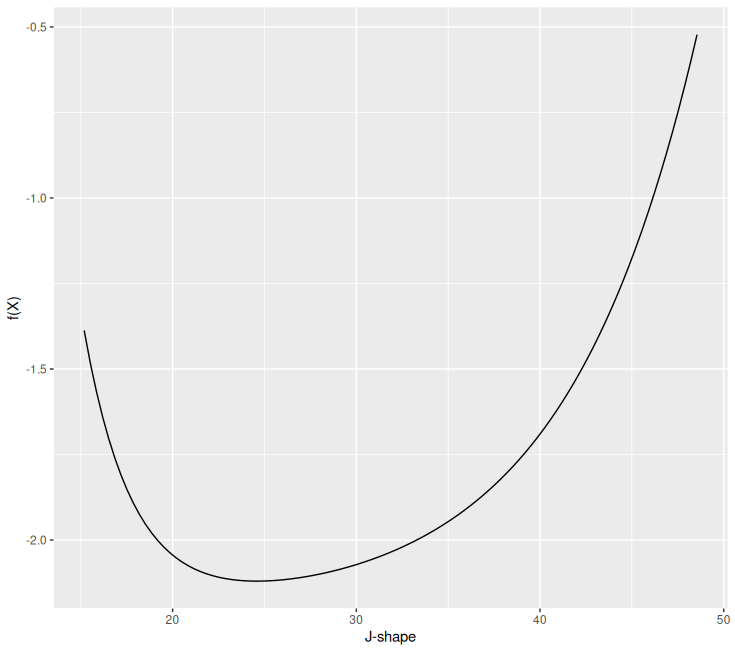}
\end{minipage}\hfill
\begin{minipage}{0.31\textwidth}
  \centering
  \includegraphics[width=\textwidth]{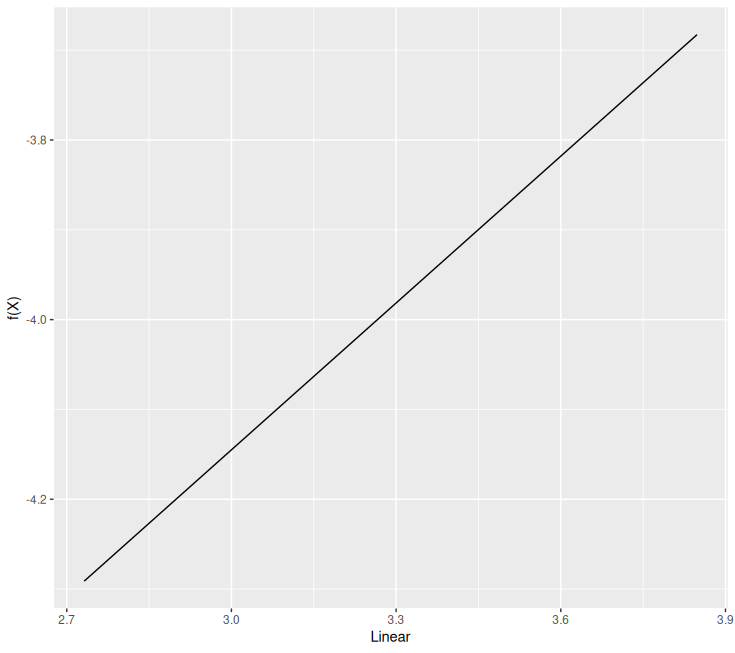}
\end{minipage}\hfill
\begin{minipage}{0.31\textwidth}
  \centering
  \includegraphics[width=\textwidth]{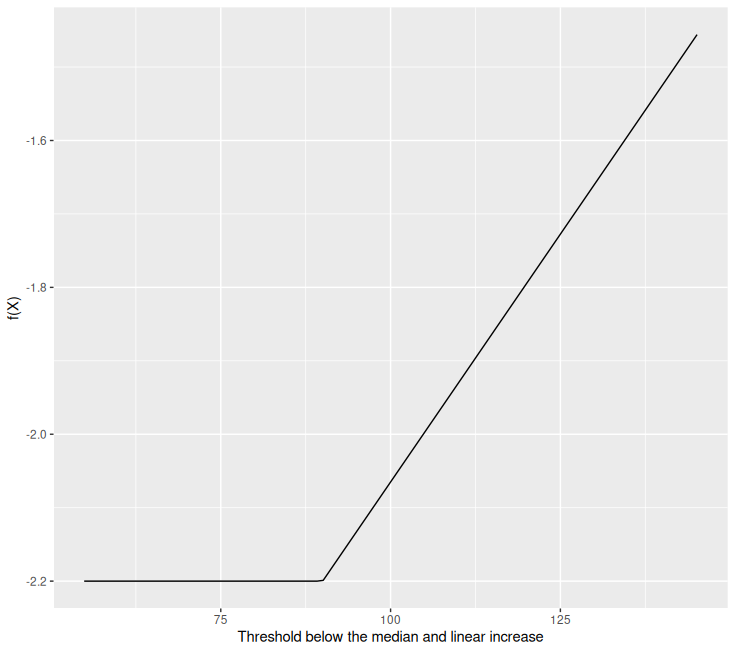}
\end{minipage}

\vspace{12pt} 

\begin{minipage}{0.31\textwidth}
  \centering
  \includegraphics[width=\textwidth]{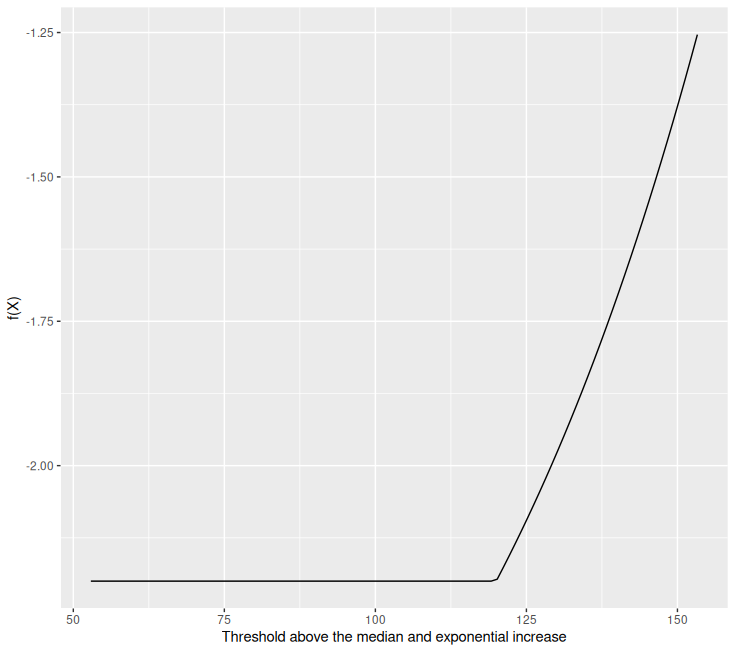}
\end{minipage}\quad
\begin{minipage}{0.31\textwidth}
  \centering
  \includegraphics[width=\textwidth]{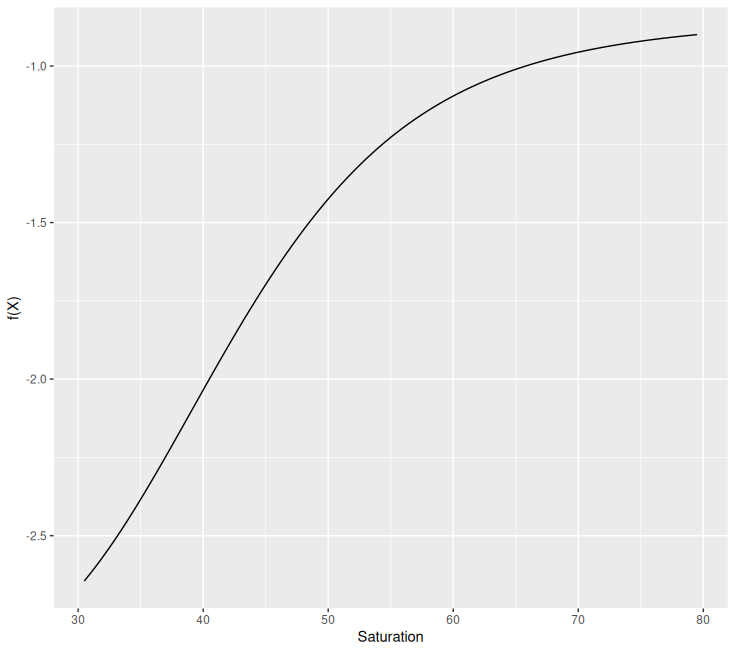}
\end{minipage}
\caption{Illustration of the five different functions $f(X)$
corresponding to the five dataset simulation scenarios.
\textbf{Scenario~1}: J-shape. Inspired by the relationship between
body mass index (BMI) and all-cause mortality, showing elevated risks
at both very low and very high covariate values.
\textbf{Scenario~2}: Linear. Reflects a constant proportional
association, similar to the relationship reported between dietary fat
intake and breast cancer incidence.
\textbf{Scenario~3}: Threshold with change-point below median and
linear increase. Models a scenario where a risk increase occurs only
after exposure exceeds a specific threshold, common in air pollution
and mortality studies.
\textbf{Scenario~4}: Threshold with change-point above median and
exponential increase. A threshold model assuming a more rapid,
exponential increase in risk once the high-exposure threshold is
surpassed.
\textbf{Scenario~5}: Saturation. Inspired by the Hill equation in
pharmacology, representing a dose-response relationship where the risk
of an effect plateaus at high exposure levels.\label{fig:functions}}
\end{figure*}

The initial stage of the simulation study, Stage 1, involved the generation of 5 distinct pilot datasets that we call scenarios. Each scenario corresponded to one of five pre-defined functional forms $f\big(X\big)$ designed to represent a variety of relationships commonly encountered in epidemiological research: a \textit{J-shape}, a \textit{linear trend}, a \textit{threshold effect with a change-point below the median} of $X$ and a linear increase beyond the threshold, \textit{ a threshold effect with a change-point above the median} of $X$ and an exponential increase beyond the threshold, and a \textit{saturation effect}. The functional forms are illustrated in Figure~\ref{fig:functions}. The mathematical details of these scenarios were: 

\paragraph{Scenario 1: J-shape} This scenario was inspired by the J-shaped relationship between body mass index (BMI) and all-cause mortality. The functional form was:\\
\begin{equation*}
 f\big(X\big) = -2.202 + 268 \exp(-0.383X) + 0.00197\exp(0.139X), \quad \text{where} \quad \ln(X) \sim \mathcal{N}\big(3.3,0.25^2\big).
\end{equation*}

\paragraph{Scenario 2: Linear} The linear association reported, for example, between log-dietary fat intake and breast cancer incidence. The functional form was: \\
\begin{equation*}
 f\big(X\big) = -5.78 + 0.545X, \quad \text{where} \quad X\sim \mathcal{N}\big(3.29,0.24^2\big).
\end{equation*}

\paragraph{Scenario 3: Threshold with change-point below median and linear increase} This scenario was based on models for air pollution and mortality, postulating a linear increase in risk only after exposure exceeds a certain threshold. The change-point was set below the median of $X$ and the functional form was: \\
\begin{equation*}
f\big(X\big) = -2.2 + 0.0135 T\big(X-90\big), \quad\text{where} \quad T\big(W\big) = W \mathbf{1}_{W\ge 0} \quad \text{and} \quad X \sim \mathcal{N}\big(100,19.4^2\big).
\end{equation*}
 
\paragraph{Scenario 4: Threshold with change-point above median and exponential increase} This scenario was also based on air pollution models but assumed an exponential increase in risk above the threshold, with the change-point set above the median of $X$. The functional form was:\\
\begin{equation*}
f\big(X\big) = -3.2 + \exp\big(0.02 T(X-120)\big), \quad\text{where} \quad \ln\big(X\big)\sim \mathcal{N}\big(4.5,0.23^2\big).
\end{equation*}

\paragraph{Scenario 5: Saturation} This scenario was inspired by the Hill equation used in pharmacology to model a drug's dose-response relationship, where the risk of a side effect saturates at high doses. The functional form was: \\
\begin{equation*}
f\big(X\big) = \ln\Big(\frac{3X^6}{7} + \frac{50^6}{19}\Big) - \ln\big(50^6 + X^6\big), \quad \text{where} \quad X \sim \text{Unif}(30,80).
\end{equation*}

For these Stage 1 datasets, the main study sample size $N$ was fixed at 15000, and the replication substudy size $n$ was 250. The measurement error variance $\sigma^2_U$ was set to $0.5 \text{Var}(X)$ representing moderate size error, and the error term $U_i$ was drawn from a Normal distribution. Consistent with the assumed 1:4 case-control study design, the proportion of $Y_i = 1$ outcomes was maintained at exactly 20\% across all datasets by stratified random selection from a larger sample. 

\subsubsection{Simulation Scenarios and Parameters: Stage 2 and Stage 3}

Stage 2 of the study greatly expanded the range of simulation scenarios. Building upon the 5 scenarios from Stage 1, 15 new combinations of simulation parameters were created for each scenario, resulting in 150 datasets. The parameters varied were main study sample size $N=(15000, 30000)$ and later expanded to smaller samples 5000 and 2000, replication sub-study sizes $n=(250, 750)$, measurement error variance $\sigma_U^2 = 0.5\text{Var}\big(X\big)$ or $\text{Var}\big(X\big)$, and the distribution of the measurement error $U_i$ (Normal or a shifted-Gamma distribution with shape parameter 3, scaled to have the target variance). All combinations of parameters were included except the combination used in Stage 1. The 15 combinations of simulation parameters generalised in Table~\ref{tab:1} were identical for the 5 scenarios.

\begin{table}[!ht]
    \centering
    \begin{tabular}{|l|c|c|c|c|}
    \hline
         & \textbf{nstudy} & \textbf{vratio} & \textbf{errdist} & \textbf{nrep} \\ 
        \hline
        comb1  & 30000/5000 & 0.5/0.5 & 0/0 & 250/250 \\ 
        comb2  & 15000/2000 & 1/0.5   & 0/0 & 250/250 \\ 
        comb3  & 30000/5000 & 1/1     & 0/0 & 250/250 \\ 
        comb4  & 15000/2000 & 0.5/1   & 1/0 & 250/250 \\ 
        comb5  & 30000/5000 & 0.5/0.5 & 1/1 & 250/250 \\ 
        comb6  & 15000/2000 & 1/0.5   & 1/1 & 250/250 \\ 
        comb7  & 30000/5000 & 1/1     & 1/1 & 250/250 \\ 
        comb8  & 15000/2000 & 0.5/1   & 0/1 & 750/250 \\ 
        comb9  & 30000/5000 & 0.5/0.5 & 0/0 & 750/750 \\ 
        comb10 & 15000/2000 & 1/0.5   & 0/0 & 750/750 \\ 
        comb11 & 30000/5000 & 1/1     & 0/0 & 750/750 \\ 
        comb12 & 15000/2000 & 0.5/1   & 1/0 & 750/750 \\ 
        comb13 & 30000/5000 & 0.5/0.5 & 1/1 & 750/750 \\ 
        comb14 & 15000/2000 & 1/0.5   & 1/1 & 750/750 \\ 
        comb15 & 30000/5000 & 1/1     & 1/1 & 750/750 \\ 
        \hline
    \end{tabular}
    \caption{\label{tab:1} Simulation parameters for each dataset scenario in Stages 2 and 3. \textbf{nstudy} corresponds to the sample size of main study. \textbf{vratio} is the ratio of variance of measurement error to variance of true exposure $X$. \textbf{errdist} corresponds to distribution of the measurement error $0=$normal, $1=$shifted-gamma. \textbf{nrep} corresponds to the sample size of replication sub-study.}
\end{table}

 In Stage 3, each of the 150 Stage 2 scenario-parameter combinations was simulated 10 times to provide repeat observations. The analysis reported herein is based on these 1500 Stage 3 datasets.

\subsection{Estimands}

The estimands of interest were the values of the true function $f'(X)$ at the pre-selected series of $X$ values detailed in subsection~\ref{part:metrics}. The adjustment was made to reflect the 1:4 nested case control ratio in the data generalised, and was given by $f'(X) = f(X) + \log(0.25*(1-p)/p)$, where $p = P(Y=1)$ in the population.

\subsection{Methods}
\label{sec:methods}

This section presents the four methods of adjusting the estimate of $f(X)$ for measurement error in the covariate $X$, each applied to four flexible modelling approaches. The first two methods, RC and Multiple MI, share the principle of first substituting an imputed $X$ in place of the unobserved $X$ and then using this imputed X in the regression model to estimate the function $f(X)$. The SIMEX method and Bayesian methods are based on completely different principles, as explained later in this Section. 

\subsubsection{Regression Calibration}

RC is based on the idea of replacing the true unobserved covariate $X$ with its imputed value, that is, an estimate of the conditional expectation $\mathbb{E}\big(X\mid X^\ast \big)$, as proposed by~\cite{prentice82}. Here we are assuming that $X$ is the only covariate in the regression model for $Y$. When, as in our study, the measurement error in $X^\ast$ is classical and there are replicate measurements of $X^\ast$ in a random subset, $\mathbb{E}\big(X\mid X^\ast \big)$ can be consistently estimated using methods described in the literature (\cite[subsection 6.1]{keogh20} and \cite[chapter 4]{carroll06}).

\subsubsection{Multiple Imputation}

When $(X^\ast, Y)$ are observed in the main study, but in a random subset of participants $X$ is also observed, then estimating relationships between $X$ and $Y$ can be formulated as a missing data problem~\cite{little19}. In using multiple imputation, the central idea is to perform a number of imputations of $X$ based on a model for the conditional distribution of $X\mid X^\ast, Y$ (more details in ~\cite[subsection 2.4]{shaw20}). When there is CME and a replication substudy of $X^\ast$ is available (instead of a validation substudy with $X$ observed), the principle of MI methods has been adapted in~\cite{keogh14, bartlett15}.

\subsubsection{Simulation Extrapolation}

The SIMEX method introduced by~\cite{cook94}, studied and extended in~\cite{carroll96}, is based on artificially increasing the variance of the measurement error $U$ by simulation, obtaining a series of estimates under this increasing error and then extrapolating back to obtain the estimate under no measurement error. Increasing the error is called the simulation step. Under the classical error model, this step consists of generating, for an increasing sequence of positive constants $s_m$ , $0 \leq s_1 \leq \ldots \leq s_M, m = 1, \ldots, M$, independent copies of $X^\ast_{i}$ according to $X^\ast_{i,b}(s_m) = X^\ast_i + \sqrt{s_m \cdot \text{Var}(U)}\cdot V_{i,b}, i=1, \ldots, n, b=1, \ldots, B$ , where $V_{i,b}$ is a standard Gaussian random variable. For each $b$ and $s_m$ , $X^\ast_{i,b}(s_m)$ are then used to obtain an estimate, $\widehat{\beta}^\ast_b(s_m)$, say of a regression parameter that links $\operatorname{logit}\!\big( \mathbb{P}(Y =1)\big)$ to $X^\ast_{i,b}(s_m)$. The objective of the simulation steps is to produce a series of pairs $(s_1 ,\widehat{\beta}^\ast(s_1)),\ldots ,(s_m ,\widehat{\beta}^\ast(s_M))$ where $\widehat{\beta}^\ast(s_m)$ is an empirical mean of the $B$ realisations $\widehat{\beta}^\ast_b (s_m)$.

Noting that the variance of the measurement error in $X^*_{i,b}(s_m)$ is given by $\text{Var}\big(X^\ast_s - X\big) = (1 + s) \text{Var}\big(U\big)$, a second step called the extrapolation step uses the pairs $\big(s_1, \widehat{\beta}^\ast(s_1)\big), \ldots, \big(s_m, \widehat{\beta}^\ast(s_m)\big)$ to fit a quadratic regression function $\widehat{\beta}^\ast = g\big(s, \Gamma\big)$ to model the estimates as a function of $s_m$ and obtain the parameter estimate $\widehat{\beta}_{\text{SIMEX}}= g\big(-1, \Gamma\big)$ by extrapolation of the function $g$ to $s=-1$ (see~\cite[chapter 5]{carroll06} for more details). 

Two implementations of SIMEX were used to estimate $\widehat{f}(X)$ depending on the statistical target of the extrapolation. The first implementation, pointwise SIMEX, involves applying the extrapolation step to the predicted value of the function at each specific evaluation point $x$ independently. For each $x$ in the evaluation range, the extrapolation function is fitted to the pairs $(s_m, \widehat{f}^*_{s_m}(x))$ generalised during the simulation phase to calculate the corrected value at $s = -1$. The second approach follows the standard SIMEX algorithm (hereafter referred to as coefficient-based SIMEX to distinguish it from the pointwise variant), where the estimated coefficients of the spline or FP basis are extrapolated directly. These corrected coefficients are then used to reconstruct the final functional form. Coefficient-based SIMEX could not be used with P-splines, as the number of coefficients forming the spline basis could vary from one iteration to another. 

\subsubsection{Bayesian methods}

As for any Bayesian analysis, one first formulates a joint distribution of $\big(Y, X, X^\ast\big)$ of observed and unobserved data. This joint distribution is then decomposed into a product of conditional distributions assuming non-differential measurement error, obtaining $f\big(y, x, x^\ast\big) = f(x^\ast \mid x) f(y\mid x) f(x)$ (see~\cite[chapter 4]{gustafson04} and \cite{bartlett18} for more details). Bayesian methods for measurement errors consider $X$ as an unobserved quantity, like the values of the parameters that allow the previous conditional distributions to be parameterised. The objective in this framework is to sample the joint posterior distribution of $X$ and the parameters of the conditional distributions involved in the previous decomposition, given the observed data $(X^\ast, Y)$ and $(X^\ast_1, X_2^\ast, Y)$ in the replication substudy. From the posterior marginal distribution of the parameters of $f(y\mid x)$, a Bayesian estimate of the model parameters is obtained. Two posterior inferences were investigated: one inference was based on the posterior mean on the $\text{logit}\left(\mathbb{P}\big(Y=1\big)\right)$ scale (which we refer to as "Bayes logit"), and one inference was based on the \textit{logit} of the posterior mean on the $\mathbb{P}\big(Y=1\big)$ scale (which we refer to as "Bayes risk").

\subsubsection{Flexible modelling approaches to estimate $f(X)$}
\label{subsec:flex}

Each of the measurement error adjustment methods generalised above was applied to four flexible modelling techniques, as described below.
\begin{enumerate}
    \item \textbf{B-splines:} Cubic regression BS were used. A single interior knot was placed at the median of the observed $X^\ast_i$, resulting in 4 degrees of freedom (3 for the degree of polynomial + 1 for the number of interior knots)~\cite{perperoglou19}. Exterior knots were typically placed at both extremes of the range of $X_i^\ast$ (for more details see~\cite{hastie09,wood17, perperoglou19}).

    \item \textbf{P-splines:} Penalised splines, typically cubic BS with 10 interior knots placed at quantiles of $X_i^\ast$ were employed. The penalty on the differences between adjacent coefficients was optimised using criteria like generalised cross-validation (GCV) or AIC, depending on the specific method and software.
   
    \item \textbf{Fractional Polynomials:} Second degree fractional polynomials (FP2) which allowed estimating non-monotone exposure-outcome relationships were investigated.
    
    \item \textbf{Natural-splines:} To avoid the erratic behaviour of polynomials near the boundaries, a natural cubic spline adds additional constraints to BS, namely that the function is linear beyond the boundary knots. These two constraints reduce the degrees of freedom compared to the BS, resulting in 2 degrees of freedom~\cite{hastie09, wood17}. For NS, knots were placed at the 5th, 50th and 95th percentiles of $X^\ast$. BS, FP and PS were investigated in a blinded manner as described above, but NS was added only at Stage 3 after the blindness was removed.
\end{enumerate}

\subsection{Performance measures}
\label{part:metrics}

To compare the efficacy of the different measurement error adjustment strategies and flexible modelling techniques, performance metric was pre-specified. The methods teams returned their estimated functional relationships, $\widehat{f\big(X\big)}$, in the form of predicted values at a series of 200 points $\{x_1, x_2, \ldots, x_{200}\}$ specified by the Data Generation and Evaluation team. The reporting ranges were specified as: $10.0-50.0$ for Dataset 1; $2.6-4.0$ for Dataset 2; $45.0-155.0$ for Dataset 3; $40.0-160.0$ for Dataset 4; and $18.0-92.0$ for Dataset 5. 
However, a subset of these points was chosen by the Data Generation and Evaluation team to evaluate the methods, covering only the central 95\% of the distribution of true covariate $X$ by defining \textit{limit points} $x_\text{low}$ and $x_\text{high}$ within which the evaluation was conducted. This ensured that comparisons were made over regions with adequate data support and avoided extrapolation into extreme tails where estimates might be unstable. These limit points were not disclosed to the Methods teams until after Stage 2.

The primary metric for assessing the accuracy of the estimated function was the log Mean Squared Error ($\log{\text{(MSE)}}$), calculated for each dataset as:

\begin{equation*}
\log{\text{(MSE)}} = \log\left\{\frac{1}{\sum_{j=1}^m \mathbb{1}\{x_\text{low} \le x_j \le x_\text{high}\}} \sum_{j=1}^m \left(\widehat{f(x_j)} - f(x_j)\right)^2 \mathbb{1}\{x_\text{low} \le x_j \le x_\text{high}\} \right\}, 
\end{equation*}
where the indicator function $\mathbb{1}\{x_\text{low} \le x_j \le x_\text{high}\}$ allows the calculation to be restricted to evaluation points falling within $\big[x_\text{low}, x_\text{high}\big]$ range. The log transformation of MSE was used to achieve a more symmetric distribution of the performance metric and stabilise its variance. Lower values of $\log{\text{(MSE)}}$ indicate a closer agreement between the estimated and true functional forms, and thus superior performance. For reporting, mean $\log{\text{(MSE)}}$ values were calculated and were exponentiated, to provide geometric mean MSEs. Geometric mean MSEs are reported averaging over the replicated simulations of all parameter combinations for each of the five $Y-X$ scenarios (300 simulations for each) and over all 1500 simulations. Secondary analysis examining interactions between methods and dataset parameters were also conducted, using linear modelling of the log MSEs. 

\section{Results}
\label{sec:results}
This section details the comparison of the performance of the different methods used for estimating the functional relationships between $X$ and $Y$ in the simulated datasets. As described in Section~\ref{sec:methods}, there are 23 analytic methods, within six families of measurement error adjustment methods: Bayes risk, Bayes logit, MI, RC, pointwise SIMEX, and coefficient-based SIMEX.

Subsection~\ref{subsec:comparison} compares the overall performance of the analytical methods averaged over all 1500 datasets and also the 300 datasets within each of the five scenarios (Y-X relationships). Subsection~\ref{subsec:performance} presents the results of the analysis of how the performance of the different methods varied with other dataset characteristics.

\subsection{Comparison of the accuracy of competing methods}
\label{subsec:comparison}
Geometric mean MSEs are presented in Table~\ref{tab:3} for the 23 methods for each scenario (first 5 columns) and for all simulated datasets combined (final column). The first four rows also present benchmark MSEs when the corresponding flexible modelling approaches were applied to the true $X$ values. Note that each cell in the first 5 columns of Table~\ref{tab:3} shows a geometric mean of 300 MSEs, while a cell in column 6 shows a geometric mean of 1500 MSEs.

\paragraph{All scenarios combined:}
Combining across all datasets, SIMEX methods achieved the best performance measures. Pointwise SIMEX was slightly superior to coefficient-based SIMEX, and results with PS, FP and NS were better than with BS. Next best to SIMEX were Bayes methods and RC when used with PS, FP or NS. MI performed less well, but the poorest results were obtained when Bayes methods were used with BS. However, the ordering of the methods differed somewhat according to the scenario, as described below.

\paragraph{Scenario 1 (J-Shape):} 
For this scenario, SIMEX was markedly better than other methods, RC was markedly poorer, and MI performed as well as Bayes methods.

\paragraph{Scenario 2 (Linear):} 
For this scenario RC using PS performed as well as SIMEX. Bayes methods and other RC methods were only a little poorer than SIMEX methods, and were actually superior to SIMEX with BS.

\paragraph{Scenario 3 (Threshold below median):} 
For this scenario, Bayes and RC were nearly as accurate as SIMEX methods, and again were superior to SIMEX with BS.

\paragraph{Scenario 4 (Threshold above median):}
In this scenario SIMEX was markedly superior to other methods, and RC performed more poorly than other methods.

\paragraph{Scenario 5 (Saturation):}
In this scenario, RC performed as well as SIMEX methods. 

\begin{table}[!ht]
\scriptsize
    \centering
    \begin{tabular}{|l|c|c|c|c|c|c|}
    \hline
    \textbf{Method} & \textbf{Scenario 1} & \textbf{Scenario 2} & \textbf{Scenario 3} & \textbf{Scenario 4} & \textbf{Scenario 5} & \textbf{Average}\\
    \hline 
    Without error PS & 5.06 (4.81, 5.33) &   0.57 (0.47, 0.69) & 1.91 (1.74, 2.11) & 2.61 (2.43, 2.81)   & 2.08 (1.89, 2.29) & 1.97 (1.88, 2.07)\\
    Without error BS & 4.24 (4.02, 4.48) &   1.84 (1.65, 2.04) & 2.26 (2.06, 2.48) & 3.04 (2.82, 3.27)   & 2.79 (2.56, 3.04) & 2.72 (2.62, 2.83)\\
    Without error FP & 11.77 (11.26, 12.31)&  1.09 (0.95, 1.25) & 1.94 (1.77, 2.13) & 3.99 (3.81, 4.19)   & 2.62 (2.44, 2.82) & 3.04 (2.93, 3.16)\\
    Without error NS & 14.56 (13.99, 15.16)&  1.01 (0.87, 1.16) & 2.01 (1.85, 2.19) & 8.17 (7.87, 8.49)   & 1.51 (1.36, 1.67) & 3.25 (3.12, 3.39)\\
    SIMEX-PS          & 5.9 (5.4, 6.4)     &   1.2 (1.0, 1.5)   & 5.5 (5.1, 5.9)     & 5.1 (4.7, 5.5)   & 14.1 (13.0, 15.3) & 4.9 (4.7, 5.2)\\
    SIMEX-FP          & 5.7 (5.3, 6.2)     &   2.4 (2.1, 2.7)   & 5.4 (5.0, 5.9)     & 4.7 (4.4, 5.1)   & 13.9 (13.0, 14.9) & 5.5 (5.2, 5.7)\\
    SIMEX-NS          & 6.0 (5.6, 6.4)     &   2.3 (2.0, 2.6)   & 5.0 (4.6, 5.4)     & 7.2 (6.9, 7.5)   & 10.3 (9.5, 11.2) & 5.5 (5.3, 5.7)\\
    SIMEX-FP-coef     & 7.4 (6.5, 8.4)     &   2.8 (2.4, 3.4)   & 5.2 (4.3, 6.2)     & 7.0 (6.3, 7.8)   & 14.5 (12.0, 17.5) & 6.4 (6.0, 6.9)\\
    SIMEX-NS-coef     & 10.4 (9.7, 11.2)   &   2.5 (2.2, 2.9)   & 4.9 (4.5, 5.4)     & 9.2 (8.7, 9.7)   & 8.7 (8.0, 9.5)     & 6.4 (6.1, 6.6)\\
    SIMEX-BS          & 8.1 (7.4, 9.0)     &   5.2 (4.7, 5.8)   & 7.6 (6.8, 8.4)     & 7.4 (6.8, 8.0)   & 13.5 (12.4, 14.8) & 8.0 (7.6, 8.3)\\ 
    SIMEX-BS-coef     & 10.0 (9.1, 11.0)   &   6.7 (6.1, 7.4)   & 9.4 (8.4, 10.4)    & 8.6 (7.8, 9.4)   & 12.9 (11.7, 14.2) & 9.3 (8.9, 9.7)\\
    Bayes-FP logit    & 34.3 (31.6, 37.1)   &   3.5 (3.0, 4.1)   & 4.9 (4.4, 5.6)     & 16.9 (15.8, 18.1)   & 17.4 (15.3, 19.7) & 11.2 (10.6, 11.8)\\
    Bayes-NS logit    & 41.6 (38.7, 44.7)   &   3.5 (3.1, 4.1)   & 5.4 (4.8, 6.0)     & 18.9 (17.7, 20.2)   & 13.2 (11.5, 15.2) & 11.4 (10.9, 12.0)\\
    RC-PS             & 85.7 (78.3, 93.7)   &   1.4 (1.1, 1.7)   & 6.1 (5.6, 6.7)     & 26.5 (24.1, 29.1)   & 10.3 (9.3, 11.5) & 11.4 (10.8, 12.1)\\
    Bayes-FP risk     & 33.4 (30.8, 36.2)   &   3.7 (3.1, 4.3)   & 5.1 (4.5, 5.8)     & 16.4 (15.4, 17.6)   & 21.8 (19.0, 25.0) & 11.7 (11.1, 12.4)\\
    Bayes-NS risk     & 40.8 (37.9, 44.0)    &   3.7 (3.1, 4.2)   & 5.5 (4.9, 6.1)     & 18.5 (17.3, 19.8)   & 14.7 (12.7, 16.9) & 11.7 (11.1, 12.3)\\
    RC-NS             & 79.7 (73.8, 86.0)   &   3.8 (3.4, 4.4)   & 6.0 (5.4, 6.7)     & 27.4 (25.3, 29.6)   & 6.1 (5.4, 6.8)     & 12.5 (12.0, 13.1)\\
    RC-FP             & 79.2 (72.6, 86.3)   &   4.1 (3.6, 4.6)   & 6.0 (5.4, 6.7)     & 25.9 (23.7, 28.3)   & 9.8 (9.1, 10.6)     & 13.8 (13.2, 14.4)\\
    Bayes-PS logit    & 36.2 (33.3, 39.4)    &   5.7 (5.0, 6.5)   & 7.4 (6.7, 8.3)     & 15.8 (14.4, 17.4)   & 24.1 (21.9, 26.6) & 14.2 (13.6, 14.9)\\
    Bayes-PS risk     & 35.8 (32.5, 39.4)   &   6.5 (5.7, 7.5)   & 8.1 (7.2, 9.0)     & 16.4 (14.8, 18.2)   & 30.0 (26.7, 33.8) & 15.6 (14.8, 16.4)\\
    MI-PS             & 33.5 (30.3, 36.9)   &   10.8 (8.7, 13.0)  & 12.7 (10.6, 15.3)  & 21.9 (19.7, 24.2)   & 28.2 (24.0, 33.2) & 19.5 (18.2, 20.9)\\
    MI-FP             & 32.7 (29.5, 36.1)   &   12.0 (9.7, 14.6)  & 13.7 (11.3, 16.6)  & 22.3 (20.1, 24.7)   & 30.5 (25.8, 36.1) & 20.6 (19.1, 22.0)\\
    MI-NS             & 39.4 (35.9, 43.3)   &   11.4 (9.3, 13.8)  & 14.1 (11.8, 16.9)  & 25.8 (23.4, 28.4)   & 27.1 (22.6, 32.5) & 21.4 (19.9, 22.9)\\
    MI-BS             & 33.6 (30.4, 37.1)   &   16.1 (9.8, 26.0)  & 17.4 (10.5, 28.7)  & 22.3 (20.0, 24.8)   & 31.5 (26.7, 37.2) & 23.1 (19.9, 26.7)\\
    RC-BS             & 87.8 (79.9, 96.5)   &   15.9 (14.2, 17.5) & 17.9 (16.1, 19.9)  & 40.2 (36.0, 44.9)   & 29.1 (25.9, 32.5) & 31.1 (29.7, 32.6)\\
    Bayes-BS logit    & 134.9 (121.0, 150.4)& 70.7 (61.3, 81.5) & 54.5 (47.8, 62.2)  & 64.0 (56.3, 72.7)   & 90.6 (81.3, 101.0) & 78.7 (74.4, 83.2)\\
    Bayes-BS risk     & 216 (186, 250)      &   206 (173, 245)    & 133 (111, 159)     & 118 (101, 137)   & 186 (163, 213)     & 167 (155, 179)\\ 
    \hline
    \end{tabular}
    \caption{\label{tab:3} Geometric mean $\text{MSEs} \times 1000$ for small and large sample sizes combined with 95\% CI based on normal assumption.}
\end{table}

\subsection{Performance analysis based on simulation parameters and methods}
\label{subsec:performance}

This subsection expands on the results shown in Subsection~\ref{subsec:comparison}. Regarding flexible modelling, after controlling for the measurement error adjustment method, the three flexible modelling methods PS, FP, and NS gave equally good results. However, unpenalised BS generalised performed worse. 

Analysing dataset characteristics revealed that, as expected, the linear relationship was the easiest to estimate, while the J-shape was the most difficult. This gradation likely reflects the challenge of correcting for attenuation; recovering a subtle, non-monotonic feature like a nadir requires higher precision than estimating a global linear slope. Larger sample sizes (main and replication) and smaller measurement errors naturally reduced MSE. Interestingly, the analysis indicated strictly lower MSE when the error distribution was shifted-gamma compared to normal. This counter-intuitive finding appears to be driven largely by extreme instability within the Bayes BS method under the normal error assumption, rather than an inherent advantage of the gamma distribution for correction methods generally.

We also observed notable interactions. Pointwise SIMEX proved to be the most robust approach, remaining optimal across all sample sizes. However, the SIMEX coefficient method became competitive at the largest sample size ($N=30,000$) and appeared resilient to larger measurement errors, suggesting its asymptotic properties hold well even under stress. Conversely, PS combined with RC performed excellently at small sample sizes but lost their competitive edge in larger samples; this suggests that while penalisation stabilises estimates when data is sparse, the associated bias may dominate the MSE as the sample size increases. The Bayesian methods using BS exhibited specific fragility, performing particularly poorly when measurement error was large.

\section{Discussion}
\label{sec:discussion}

This paper presents the results of a comprehensive, neutral comparison simulation study designed to compare measurement error adjustment methods generalised with flexible modelling methods. The challenge of handling measurement error is substantial, as unadjusted analyses can distort the true functional form, leading to biased estimates, reduced statistical power, and unreliable scientific conclusions. Our study provides a rigorous and neutral comparison of several advanced adjustment techniques combined with flexible modelling strategies across a wide range of realistic scenarios.

A striking finding from our study was how surprising the results were to the research team. Based on theoretical considerations, the widely held expectation was that Bayesian and MI methods would outperform RC, which in turn would be superior to SIMEX. However, our results showed a different hierarchy of performance. SIMEX, particularly the pointwise implementation, consistently performed best and was the most accurate approach across the five scenarios. Following SIMEX, Bayes methods and RC when used with PS, FP or NS. MI performed less well, but the poorest results were obtained when Bayes methods were used with BS, consistently exhibiting the highest mean squared error. A plausible explanation is that BS lack the constraints imposed by the other methods (such as the linearity constraints of NS or the roughness penalties of PS), making them more susceptible to variance inflation and instability, particularly in the tails of the distribution when correcting for measurement error. The blinded nature of the study’s initial stages was crucial in this regard; it provided an unbiased environment that lent significant credibility to these unexpected findings and helped convince any sceptics within the group.

While a clear explanation for these performance differences is currently under investigation, we suspect that SIMEX may be more robust in the complex, non-linear models assessed in this project. To fully deconstruct these performance differences, further forensic research is required. Future work should investigate why SIMEX outperformed the likelihood-based methods in this specific setting, potentially using leave-one-out influence analyses to identify whether specific data points disproportionately affect the Bayesian posterior or MI estimates. Additionally, simulation stress-tests reducing the sample size to a ‘breaking point’ could help determine the stability limits of the extrapolation step in SIMEX versus the convergence properties of the MCMC in the Bayesian approach.

Related to our study design, it is important to contextualise how the flexible modelling approaches FP and splines were implemented compared to routine practice. In the FP approach, we decided to restrict analyses to the class of FP2 functions, which has 36 pairs of power terms $p_1$ and $p_2$, and the best fitting FP2 function was selected. We did not allow a larger class of functions, nor did we use the full Function Selection Procedure (FSP) often employed in practice with one variable. FSP is highly relevant in the multivariable context, but it utilises a series of tests which require fixing a p-value or an information criterion~\cite{royston08}. However, we reasoned that if the true function is linear, the best FP2 function is usually not far from linearity. Similarly, for the spline approaches (BS, PS, and NS), we employed standardised knot placement strategies (typically based on quantiles) and fixed degrees of freedom (for BS and NS) or fixed penalty optimisation procedure (for PS) to ensure comparability across the simulation batches. In clinical practice, knot selection and penalty optimisation might be more interactive or rely on different optimisation criteria or visual inspection, which could introduce further variability into the estimates.

A subtle but critical aspect of this study is the potential for ‘analyst variability’. While we compared methods, these were implemented by specific expert teams within the STRATOS initiative. It is possible that different team members, were they to implement the methods alone, might derive slightly different results. For example, in a Bayesian analysis, the choice of prior distributions or the specific tuning of the MCMC sampler can influence the posterior; in SIMEX, the choice of the grid for the simulation step or the extrapolant function can alter the final estimate. While our study protocol harmonised many of these choices to ensure a fair comparison, readers should be aware that the \textit{performance of a method} is often inextricably linked to the implementation decisions made by the analyst.

Table 2 reveals signs of a hierarchy in performance, accompanied by notable variability across the simulation scenario. The SIMEX family, particularly when coupled with PS or FP, appears to perform well in most scenarios and frequently occupies the highest rankings. In contrast, the relative standing of RC (RC) and certain Bayesian methods shows more frequent fluctuations; for instance, RC-PS ranks highly in the linear and saturation scenarios but occupies a lower position in the J-shape and threshold-above-median scenarios. The MI family tends to cluster in the lower-middle portion of the rankings, exhibiting less scenario-specific variation than the calibration methods. Finally, there is a consistent trend at the bottom of the rankings, where Bayes-BS (logit and risk) remains the poorest performing approach across all scenarios, which reinforcing the view that unconstrained splines are ill-suited for Bayesian correction without strong priors to control shape.
 
Although this study was extensive, it was intentionally focused on the core problem of a single covariate measured with error, and several important questions remain. Future work should expand these scenarios to be more realistic by, for instance, including several other precisely measured or error-prone covariates in the model, as is common in observational studies. Furthermore, this work assumed the error-prone covariate was of primary interest; the complex interplay between measurement error correction and simultaneous variable selection remains a critical area for investigation. Our evaluation also focused on the accuracy of the estimated function, but future stages of this project plan to understand the variances of the estimates and the impact of these methods on statistical power. Finally, the current analysis focused on relatively large sample sizes. A next step would be to investigate performance in smaller sample settings, which occur in some studies.

This work directly supports the STRATOS initiative’s goal of increasing awareness and removing barriers to the use of appropriate statistical methods~\cite{sauerbrei14}. It highlights that adjusting for measurement error when modelling non-linear relationships is not only necessary but also feasible. The choice of adjustment method can have a profound impact on the results, and researchers should be aware that no single technique is universally dominant. Crucially, this study reinforces the importance of sensitivity analyses to assess the robustness of findings. In conclusion, this neutral comparison study provides novel evidence to guide researchers, demonstrates the capabilities and potential pitfalls of current methods, and charts a clear course for future methodological research in this critical area.


\end{document}